\documentclass[fleqn,twoside,12pt]{article}
\usepackage{amsmath,amssymb,amsthm}
\usepackage{bbold}
\usepackage[dvipsnames]{xcolor}
\usepackage{comment}
\usepackage[shortlabels]{enumitem}
  \setlist[enumerate,1]{leftmargin=15pt}
  \setlist[itemize,1]{leftmargin=12pt}
  \setlist[description,1]{leftmargin=10pt}
\usepackage[T1]{fontenc}\usepackage{lmodern}
\usepackage[text={6in,9.5in},centering]{geometry}
\usepackage{graphicx}
\usepackage{microtype} 
\usepackage{qcircuit}
  \newcommand{\ew}[1][-1]{\ar @{} [0,#1]}
  \newcommand{\emeasure}[1]{*+[F-:<.9em>]{#1} \ew}
  
\usepackage{url}
\usepackage[bookmarks]{hyperref}
\newtheorem{theorem}{Theorem}[section]

\theoremstyle{definition}

\newtheorem{definition}[theorem]{Definition}
\newtheorem{example}{Example}[section]

\newtheorem{proviso}{Proviso}
\newtheorem{remark}{Remark}[section]

\newenvironment{eatab}
 {\medskip\noindent\begin{minipage}{\textwidth}%
  \normalfont\ttfamily\small
  \begin{tabbing}mmm\=mmm\=mmm\=mmm\=mmm\=mmm\=mmm\=mmm\=mmm%
               \=mmm\=mmm\=mmm\=mmm\kill}
 {\end{tabbing}\end{minipage}\medskip}

\newenvironment{lquote}
  {\list{}{\leftmargin=1.5em\rightmargin=1em}\item[]}%
  {\endlist}

\newcommand\bra[1]{\ensuremath{\langle#1|}}

\newcommand\bu{\ensuremath{\bar u}}
\newcommand\bx{\ensuremath{\bar x}}
\newcommand\cnot{\textit{controlled-NOT\,}}
\newcommand\Co{\ensuremath{\mathbb C}}
\newcommand\cs[1]{\ensuremath{\text{CSrc}(#1)}}
\newcommand{\dg}{^\dag}

\newcommand\Else{\texttt{else}}
\newcommand\Elseif{\texttt{elseif}}
\newcommand\Entries{\ensuremath{\textrm{Entries}}}
\newcommand\Exits{\ensuremath{\textrm{Exits}}}
\newcommand\et{\ensuremath{\texttt{and}}}
\newcommand\F{\ensuremath{\mathcal F}}
\newcommand\false{\texttt{false}}
\newcommand\G[1]{\ensuremath{{\mathcal G}(#1)}}
\newcommand{\gt}[1]{\framebox[1.2\width]{#1}}

\newcommand\Id{\ensuremath{\mathbb 1}}
\newcommand\If{\texttt{if}}
\newcommand\iset[1]{\ensuremath{\langle#1\rangle}}
\newcommand\Iset[1]{\ensuremath{\big\langle#1\big\rangle}}
\newcommand\ket[1]{\ensuremath{|#1\rangle}}
\newcommand\ketbra[2]{\ensuremath{|#1\rangle\langle#2|}}
\newcommand\measurement{\ensuremath{\mathtt{measurement}}}
\newcommand\mq{\ensuremath{\text{mq}}}
\renewcommand\O{\ensuremath{\Omega}}
\newcommand\on{\ensuremath{\mathtt{on}}}
\newcommand\op{\ensuremath{\oplus}}
\newcommand\OV[1]{\ensuremath{\text{OV}(#1)}}
\newcommand\ox{\ensuremath{\otimes}}
\newcommand\PM{\ensuremath{PM}}
\newcommand\pr{\ensuremath{\parallel}}
\newcommand\qef{\hfill$\triangleleft$} 
\newcommand\QFT{\ensuremath{QFT}}
\newcommand\qstate{\ensuremath{\texttt{qstate}}}

\newcommand\random{\texttt{random}}
\newcommand\Set[1]{\ensuremath{\big\{#1\big\}}}
\newcommand\set[1]{\ensuremath{\{#1\}}}
\newcommand\SM{\ensuremath{SM}}

\newcommand\Sub[1]{\ensuremath{\text{Sub}(#1)}}
\newcommand\swap{\ensuremath{swap}}

\newcommand\Then{\texttt{then }}

\newcommand\true{\texttt{true}}

\newcommand\Vdots{\overset{\vdots}{\phantom{o}}}

\newcommand\wq{\ensuremath{\text{wq}}}

\title{Software science view on\\ quantum circuit algorithms}
\author{Yuri Gurevich and Andreas Blass\\
\normalsize University of Michigan, Ann Arbor, Michigan, USA}
\date{}

\begin{document}
\maketitle

\footnotetext{Partially supported by the US Army Research Office under W911NF-20-1-0297}
\thispagestyle{empty}

\begin{abstract}
We show that, on the abstraction level of quantum circuit diagrams, quantum circuit algorithms belong to the species of interactive sequential algorithms that we studied in earlier work.
This observation leads to a natural specification language for quantum circuit algorithms.
\end{abstract}


\section{Introduction}
\label{intro}

We claim that, on the abstraction level of widely used quantum circuit diagrams, quantum circuit algorithms, \emph{QC algorithms} in short, are interactive sequential algorithms, in fact a simple special case of those studied in \cite{G166-171}. But QC algorithms are approximate, parallel, and randomized.
How can we claim that they are sequential?
We will address these objections shortly.
First we discuss what we mean by QC algorithms.

\subsection{What are QC algorithms?}
\label{sb:what}

Recently, in \cite{G250}, we defined syntax and semantics for quantum circuits with classical channels and for algorithms over such circuits.
In the present introduction we recall some of those definitions.
We also use some of the notions from \cite{G250} intuitively but we recall those definitions in \S\ref{s:qpre} to make this paper more self-contained.

To simplify the presentation, we restrict attention to qubit-based circuits with wires.
Wires are also known in the literature as timelines or registers.
If a circuit has $w$ input qubits (all input qubits, including ancillas, if any), we say that $w$ is the \emph{width} of the circuit; the state space for such a circuit is $(\Co^2)^{\ox w}$.

A nonempty set $B$ of gates of a circuit $C$ forms a \emph{bout} if no $B$ gate is a prerequisite for another $B$ gate.
Thus, the gates of a bout may be executed in parallel in a single step.
A \emph{schedule} for $C$ is a sequence $B_1, B_2, \dots, B_t$ of disjoint bouts such that every gate is in one of the bouts and all its prerequisites are in earlier bouts%
\footnote{More liberal schedules are discussed in Remark~\ref{r:par}.}.

A \emph{quantum circuit algorithm} $A$ is given by a circuit $C$ and a schedule $B_1, B_2, \dots, B_t$ for $C$.
The algorithm $A$ works in $t$ steps.
In step~1, the $B_1$ gates fire; in step~2, the $B_2$ gates fire; and so on.
It was shown in \cite{G250} that the input-output behavior of $A$ depends only on $C$, not on the schedule.

\subsection{Imprecision, parallelism, and randomness}
\label{sb:ipr}

We return to the objections, raised above, to our claim that QC algorithms are sequential algorithms interacting with the environment.
In the course of the discussion, we address two common misconceptions about sequential algorithms.

\paragraph{\tt Imprecision.}
Indeed, the inputs and outputs of a QC algorithm are in general approximate, and so are the operations of its gates.
But the approximation aspect, so important in experimental physics, is completely ignored by the quantum circuit model which abstracts from the provenance of inputs and inaccuracy of gate operations.
A QC algorithm with schedule $B_1, B_2, \dots, B_t$ computes in $t$ discrete steps where each step is a well-defined (at run time) transformation of the state space of the algorithm.

\paragraph{\tt Parallelism.}

One common misconception is that a sequential algorithm cannot perform several distinct actions in parallel, within one step.
In fact, it can.
For example, a Turing machine may change the control  state, write a symbol in the currently observed tape cell, and move the read/write head.
But the parallelism of a sequential algorithm $A$ is bounded by a fixed number $n$, independent of input, so that $A$ can perform at most $n$ actions in parallel.
The parallelism of a QC algorithm over a quantum circuit is bounded by the width of the circuit.

In this paper, we consider only single-circuit algorithms.
If we were to consider a (uniformly) parametrized family of QC algorithms, there would in general be no fixed bound on parallelism;
bounds for individual circuits would depend on the parameters.
There is an analogous situation with parametrized families of classical Boolean circuits.

One might also consider such a family to be a single algorithm.
But we find the complexity-theoretical definition of families unsatisfactory for programming/specification purposes \cite[\S1]{G251}.
In this connection, we hope to find a sufficiently general (at least for practical purposes) definition of parametric programs and specifications that does not require an auxiliary classical algorithm constructing quantum circuits.
We intend to address such a generalization of QC algorithms elsewhere.

\paragraph{\tt Randomness.}

A QC algorithm with measurement gates relies on its environment for performing quantum measurements.
In particular, the algorithm itself does not (and cannot) choose the
random classical outcomes of measurements.
This is done by nature.
In that sense, nature is a part of the environment of a QC algorithm.

A second common misconception is that sequential algorithms do not interact with the environment, except possibly for accepting input and emitting output.
In reality, sequential algorithms, especially those written in  high-level programming languages, depend on the environment, specifically the operating system, for various services, e.g., for providing (pseudo) random numbers.

This completes, for the time being, our discussion on the sense in which QC algorithms are sequential.
We will return to the randomness issue in \S\ref{s:intrastep}.

\subsection{Related work}

The most closely related body of work is that on abstract state machines, though we do not presume any familiarity with that body of work.

Below ``abstract state machine'' is abbreviated to ``ASM.''
The ASM thesis says that every algorithm, of any kind, is step-for-step simulated, at its natural level of abstraction, by an appropriate ASM \cite{G103}.
In \cite{G141}, the thesis was derived from first principles for sequential algorithms:
\begin{itemize}[itemsep=-2pt,topsep=3pt]
\item sequential algorithms were axiomatized by means of three postulates,
\item a \emph{sequential ASM language} was defined (the algorithms programmed in this language are called \emph{sequential ASMs}), and
\item it was proved that, for every sequential algorithm $A$, there is a sequential ASM that step-for-step simulates $A$.
\end{itemize}
Later the ASM thesis was derived from first principles for parallel algorithms in \cite{G157}, and for interactive sequential algorithms in \cite{G166-171} and \cite{G176&182}.

In article \cite{GN} Erich Gr\"adel and Antje Nowack pioneered quantum ASM research:
``We first show that, and how, a general model of quantum algorithms
(based on the combination of a classical computer with quantum circuits) can be modeled by abstract state machines \dots
we then formulate general postulates for quantum algorithms and show that every algorithm satisfying these postulates is simulated by an appropriate ASM.''
Contrary to the previous cases, however, these theoretical results did not give rise to applications, primarily because the quantum ASM programs of \cite{GN} work with orthonormal bases of the quantum state spaces and thus are exponential in the number of qubits involved.

Here we show that QC algorithms form a simple subspecies of the so-called \emph{ordinary interactive sequential algorithms} studied in \cite{G166-171}.
We introduce a language for specifying arbitrary QC algorithms.
We call this specification language the \emph{ASM language for QC algorithms}, in short \emph{QC-ASM}.

Now we turn attention to a different body of work.
There is a large and growing number of quantum programming languages and toolkits\footnote{The first author was a member of Microsoft's Quantum Architecture and Computing group when it started to develop Q\#.}.
A succinct survey of quantum languages is found in \cite{Rand}.
Of course, the present paper proposes a specification language, and specification languages are quite different from bona fide programming languages;
specification languages are more theoretical and may not have efficient real-world implementations.
Still these two kinds of languages are somewhat related, so it makes sense to discuss some programming languages here, though
we limit this discussion to those programming languages that seem most closely related to specification languages, namely Quipper (and its variants) and Qwire.

Peter Selinger and various coauthors have developed a quantum programming language, Quipper, and variants including Proto-Quipper-M and Proto-Quipper-Dyn \cite{Green+4, Green+4b, Selinger, Smith+3, Valiron+4}.
Unlike our work in the present paper, Quipper and its variants are based on the viewpoint that quantum computations are subroutines, called when needed by a classical computer.
Thus, for example, control operations are usually the responsibility of the classical computation; only in Proto-Quipper-Dyn \cite{Fu+3b} are quantum phenomena allowed to participate in control (via so-called dynamic lifting, similar to our classical channels).

Another consequence of the Quipper viewpoint is that the language is designed to operate smoothly with a classical ``host'' programming language. The host language chosen for this purpose is Haskell.  Accordingly, Quipper is, like Haskell, a functional programming language, in contrast to the imperative nature of our specification language QC-ASM.
Like many functional programming languages, Quipper is closely related to (and in some aspects inspired by) category-theoretic models \cite{Rios+1}. A recent extension \cite{Fu+3} incorporates dependent types.

There are also some lesser differences between Quipper and QC-ASM. One is that Quipper explicitly distinguishes ancillas from the meaningful qubits; it explicitly marks not only the introduction of each  ancilla but also the end of the scope in which it is used. Another difference is that measurements in Quipper are merely projective measurements with respect to the computational basis.

The programming language Qwire, introduced in \cite{Paykin+2}, is, like Quipper, an embedded language, i.e., it is to be used by a classical host to run quantum subroutines. But no particular host language is specified; Qwire is intended to operate with any of a wide spectrum of classical programming languages.

Qwire uses a combination of ordinary and linear type structures. Classical information is handled by traditional types, but quantum information has a type structure that is linear, in the sense of being based on Girard's linear logic \cite{Girard}. Linear type structure requires each qubit of quantum information to be used exactly once; it thereby enforces the no-cloning property of quantum physics. It also means that a qubit can be discarded only by measuring it, thereby turning it into classical information, and then discarding that classical bit.

Qwire, like our specification language QC-ASM, describes quantum computations in terms of wires carrying qubits.
The gates in Qwire are unitary operations on a single wire and controlled-unitary operations on a pair of wires.
Measurements are projective measurements of a single qubit in the computational basis.
QC-ASM allows far more general gates and measurements, even non-projective measurements.
It is known that the limited facilities of Qwire can be combined, also using ancillas, to simulate our more general measurements.
Our greater generality makes it possible to work at a higher level of abstraction, which is important in specifications.

The paper \cite{Paykin+2} provides both operational and denotational semantics for Qwire. The denotational semantics works with mixed states rather than pure ones.

Finally we note that Quipper and Qwire include facilities to deal with uniform circuit families rather than only single circuits.
According to the standard complexity-theoretic definition, such a circuit family is constructed by a logspace bounded Turing machine.
And indeed, Quipper and Qwire regard quantum circuits as being constructed by a classical sequential algorithm run by the classical host.
Any such classical sequential algorithm can be programmed in the sequential ASM language which is a part of QC-ASM.
But, as we mentioned in \S\ref{sb:ipr}, we find the complexity-theoretic definition of uniform circuit families unsatisfactory.

\subsection{This paper}

After two preliminary sections, \S\ref{s:qpre} and \S\ref{s:lpre}, we illustrate, in \S\ref{s:lang}, our specification language QC-ASM and show how QC algorithms can be specified in that language.
The abstraction level of QC-ASM is close to that of typical quantum circuit diagrams.
This is illustrated in \S\ref{s:examples}.

Our specifications (\emph{specs} in short) are on the abstraction level of quantum circuit diagrams and closely resemble the diagrams.
But they have well defined semantics and avoid other limitations of diagrams.
These advantages are visible already in the small examples in \S\ref{s:examples}.
There is no need to rely on intuition for the meaning of ellipses ``$\dots$'' or ``\,$\vdots$\,''.
There is no need for ad hoc notations like ``Repeat \dots\ times.''
In the Quantum Fourier Transform example, the parallel \swap\ gates were omitted from the diagram because they are hard to draw neatly.
There is obviously no difficulty including these parallel swaps in the QC-ASM spec.
Of course, the advantages of QC-ASM specs over diagrams would be even more obvious in larger examples.

Sections up to and including \S\ref{s:examples} are intended to show the reader how to use our language for specifying QC algorithms.
The next two sections are more theoretical and put this paper in the framework of abstract state machine theory.

In \S\ref{s:intrastep}, we show that QC algorithms form a simple subspecies of the interactive sequential algorithms introduced and studied in \cite{G166-171}.
Interactive sequential algorithms are intra-step interactive.
In other words, they are able to interact with the environment during a step.
Traditional sequential algorithms also interact with the environment but only between steps, not during a step.
In \S\ref{s:interstep}, we show that, by slightly lowering the level of abstraction, we can view QC algorithms as traditional sequential algorithms.
In \S\ref{s:langb}, we give formal syntax and semantics for the QC-ASM language.

The future work is related to ensuring that quantum programs run as intended.
In general, we tend to view formal verification as a form of testing.
But how do you test quantum programs?
Normally a classical computation can be stopped at any point and its state can be examined. This is not the case with quantum computations. Does this give a greater role for formal verification?

As illustrated by Example~\ref{e:cnot}, quantum programs tend to have many computation runs.
In that example, we used symbolic computation-run tracking.
How useful this approach?
Can it be automated?
These are some of the questions to consider.

We use \cite{NC} as our main reference on quantum circuits and quantum circuit algorithms.
The talk \cite{GErichFest} was a precursor of the present paper.

\section{Quantum preliminaries}\label{s:qpre}

For the reader's convenience, we recall here some relevant information.

\begin{proviso}
By default, Hilbert spaces are finite dimensional, and quantum systems have finite dimensional state spaces.
\end{proviso}

\subsection{Quantum circuits with classical channels}
\label{sb:qc}

We recall some definitions from \cite{G250} in a form appropriate for our goals in this paper.

A quantum gate computes a unitary operator or a quantum measurement.
View a unitary $U$ as a degenerate measurement with a single classical output and a single linear operator $U$.
Thus every quantum gate computes a quantum measurement.
The only purpose of imposing this uniformity is to simplify the exposition.

A quantum gate $G$ is endowed with two disjoint nonempty finite sets of the same cardinality, the set $\Entries(G)$ of the \emph{entries} of $G$ and the set $\Exits(G)$ of the \emph{exits} of $G$.
A quantum circuit comes with a one-to-one function Bind from the input nodes and gate exits to gate entries and output nodes.
If Bind$(x) = y$, we say that $x$ and $y$ are \emph{bound}.

A gate $F$ is a \emph{quantum source} for a gate $G$ if an exit of $F$ is bound to an entry of $G$, and $F$ is a \emph{classical source} for $G$ if the classical output of the measurement computed by $F$ is made available to $G$.
$F$ is a \emph{direct prerequisite} of $G$ if $F$ is a quantum or classical source for $G$.
The \emph{prerequisite} relation $\prec$ on the gates is the transitive closure of the direct prerequisite relation.

In this paper we require that a quantum gate $G$ comes with a bijection from $\Entries(G)$ to $\Exits(G)$.
Together with the Bind function, these bijections engender so-called wires, known also as timelines or registers, going from the inputs to outputs, drawn horizontally in pictures.
See examples in \S\ref{s:examples}.
We think of wires as being numbered from the top down, starting with 1.

A gate $G$ with a set \cs{G} of classical sources is assigned a finite nonempty set of \emph{potential measurements} and a \emph{selection function} $\sigma_G$ that selects, at run time, the actual measurement $\sigma_G\iset{x_F: F\in \cs{G}}$ executed by $G$; here $x_F$ is the classical output produced by gate $F$.
If \cs{G} is empty, so that $G$ has no classical sources, then there is only one measurement assigned to $G$.

\subsection{Classical channels}
\label{sb:ch}

General discourse and examples of quantum circuits in the literature seem to treat classical channels as going from one gate to one gate, not as displays.
They do not, however, explicitly prohibit displays.
As a result, the notion of ``channel'' is a bit vague.
In \cite{G250}, we adapted the one-to-one view of classical communication, according to which a gate may have more than one outgoing channel.
Now we prefer a one-to-many view because it is more efficient for programming purposes.
The mental picture of multiple outgoing channels of $G_1$ is replaced with that of a single display.
In addition we put forward the following proviso.

\begin{proviso}\label{p:ch}
Every gate $G$ has a unique outgoing channel.
\end{proviso}

Thus, upon performing its measurement, $G$ sends the classical outcome over its channel (i.e., displays the outcome).
Every gate having $G$ as a classical source is a recipient of that information.
In addition there could be outside recipients of that information, especially if the classical outcome in question is an output of the circuit.
Notice that an outside recipient may be interested only in the existence of the classical outcome as a sign that the measurement has been performed successfully.

It is not assumed that the outgoing channel has material existence or that there are interested recipients.
It may be a channel to nowhere.
The purpose of the proviso is to simplify the theory.
The channel, more exactly the associated variable, provides a
standard place to record the classical outcome of the gate.

\subsection{Measurements}
\label{sb:meas}

A general quantum measurement of a quantum system $Q$ is a finite indexed family $M = \iset{A_i: i\in I}$ of linear operators $A_i$ on the state space of $Q$ such that $\sum_i A_i\dg A_i$ is the identity operator \Id.
At each normalized quantum state $\psi$ of $Q$, the classical outcome of $M$ is a random variable taking values $i\in I$ with probabilities $p(i) = \bra\psi A_i\dg A_i \ket\psi$ \cite[\S2.2.3]{NC}.

If $I$ is a single element, the requirement $\sum_i A_i\dg A_i = \Id$ says that the unique $A_i$ is unitary.
If that unitary is the identity, we say that $M$ is the identity measurement.

As mentioned earlier, we view a unitary $U$ as a degenerate measurement with a unique classical output and a unique linear operator $U$.
While the uniformity is natural mathematically and simplifies the
exposition, a unitary is physically different from a ``genuine"
measurement because no randomness and no ``collapse of the wave function'' is involved.
We will return to this point in \S\ref{s:intrastep}.

\section{Logic preliminaries}\label{s:lpre}

\subsection{Traditional paradigm}

\begin{definition}
A \emph{(traditional many-sorted) vocabulary} is a finite collection of \emph{sort symbols}, including symbol \texttt{Bool}, and \emph{function symbols}, including symbols \true, \false, the standard propositional connectives, and the equality sign.
Each function symbol $f$ is assigned a nonnegative integer, the \emph{arity} of $f$.
Every function symbol $f$ of arity $r$ is assigned a \emph{type}
$S_1 \times \cdots \times S_r \to S_0$
where $S_0, \dots, S_r$ are sort symbols.
If $f$ is nullary, i.e.\ if $r=0$, we say that $f$ is of type $S_0$. \qef
\end{definition}

\begin{remark}
One may wonder what is the difference between sorts and types.
Sorts are a special case of types. Sorts are given by the vocabulary; you don't construct new sorts.
Types, in contrast, can be constructed from sorts by operations like $\times$ and $\to$ used above. \qef
\end{remark}

\begin{definition}\label{d:struc}
A {\em structure} $X$ of vocabulary $V$ provides \emph{denotations} for the $V$ symbols.
Each sort symbol $S$ denotes a nonempty set.
A function symbol $f$ of type $S_1 \times \cdots \times S_r \to S_0$ denotes a function $f$ from $S_1 \times \cdots \times S_r$ to $S_0$, a \emph{basic function} of $X$;
if $r=0$ then the denotation is just an element of $S_0$.

\texttt{Bool} consists of two distinct elements, (the denotations of) \true\ and \false, and the propositional connectives have their usual denotations.
Also the equality sign works as usual on all elements regardless of their sorts. \qef
\end{definition}

\begin{remark}[Names and denotations]
Syntactic objects often denote semantical objects.
For example, in Definition~\ref{d:struc}, the first occurrence of $f$ is syntactic, and the second is semantic; similarly for the types $S_0, S_1, \dots, S_r$.
Different conventions may be used for disambiguation, e.g.\ a basic function may be denoted $f_X$. We will use no disambiguation convention in this paper. It should be clear from the context whether a symbol means a syntactic or semantic object.
\qef
\end{remark}

Define \emph{expressions} (also known as \emph{terms}) by induction.
A nullary symbol of type $S$ is an expression of type $S$, also called an \emph{$S$-valued term}.
If $f$ is a function symbol of type $S_1 \times \cdots \times S_r \to S$ and if each $t_i$ is a term of type $S_i$, then $f(t_1,\dots,t_r)$ is a term of type $S$, an $S$-valued term.

\subsection{Computational paradigm}

The states of an abstract state machine are structures of a fixed vocabulary, but this requires enriching the notion of vocabulary.

\begin{definition}
An \emph{ASM vocabulary} is a traditional vocabulary where the function symbols are split into three categories: \emph{static, dynamic, external}.
The symbols \true, \false, the propositional connectives, and the equality sign are all static.
Nullary static symbols are \emph{constants}, and nullary dynamic symbols are \emph{programming variables}. \qef
\end{definition}

In the states of an abstract state machine, the denotations of the static symbols are given by initial states, those of the dynamic symbols may be changed by the algorithm, and those of the external symbols are controlled by the environment.

Below, by default, vocabularies are ASM vocabularies.

\smallskip
We presume that every vocabulary possesses the sort \texttt{integer} as well as the standard arithmetical operations, which are all static.
In every structure, the sort \texttt{integer} and the arithmetical operations have their standard denotations.

\section{An executable specification language for QC algorithms}
\label{s:lang}

In this and the following sections, we describe a simple language for programming QC algorithms on the level of common quantum circuit diagrams.
In this section, the programming constructs of the language are introduced by showing how to use them in programming an arbitrary QC algorithm.
A more formal description of the syntax and semantics of the language will be given in \S\ref{s:langb}.

We call this language the \emph{ASM language for QC algorithms}, in short \emph{QC-ASM language}, because it is a special form of the language for programming sequential abstract state machines, in short \emph{sequential ASMs} \cite[\S6]{G141}.
How special is the ``special form''? We address this question in \S\ref{sb:persp}.

In the ASM theory (and practice), states of an algorithm are rather comprehensive.
Intuitively, at any point during the algorithm's run, the state contains all the information (at the abstraction level of the algorithm) relevant to the future of the run except for the program of the algorithm and interventions of the environment.
In the absence of environmental interventions (at the algorithm's abstraction level), the current state and the program of the algorithm determine the future.

Such a comprehensive state of a QC algorithm determines its quantum state but not the other way round.
The quantum state tells us nothing about which gates, if any, are about to fire, what classical information has been received by those gates, and so on.
(Similarly, in the case of classical Turing machines, people often use the word ``state'' to mean just the control state, but the state in the ASM sense also includes the tape contents and the location of the read/write head.)

As in \cite{G141}, comprehensive states of a QC algorithm are structures of a fixed many-sorted vocabulary which includes (among others) sorts \texttt{bool} and \texttt{integer} with the usual logical and arithmetical operations.
But, in the case of a QC algorithm, the vocabulary also includes some quantum-specific sorts.
One such sort is \qstate\ (an allusion to ``quantum state''),
another is \measurement.

\subsection{Gates}
\label{sb:gate}

We start with a useful notation.
Consider what happens in run time when a gate performs a measurement $M$.
The action of the gate is not determined just by the measurement $M$ itself.
The action also depends on what wires are involved, and the order of wires may be relevant as well.
For example, the actions of performing \cnot\ on wires 1,2, on wires 2,3, and on wires 3,2 are all distinct.
A measurement $M$ on wires $\bu = (u_1, \dots, u_k)$ will be denoted $M(\bu)$.

Let $G$ be a gate operating on wires \bu.
First suppose that $G$ is assigned a single measurement $M$ and thus performs a fixed measurement $M(\bu)$.
Recall that, by Proviso~\ref{p:ch}, every gate has a unique outgoing channel.
Hence, we can introduce a \emph{channel variable} $y$ to represent the content of the outgoing channel, the result of the measurement.
The action of the gate is expressed by an \emph{assignment rule}
\begin{equation}\label{as1}
 y := M(\bu)\,.
\end{equation}

\begin{remark}
There is something odd about assignment \eqref{as1}.
On the right there is an expression of a measurement type but on the left an integer expression.
We could have used something like CO$(M(\bu))$ where CO alludes to ``classical outcome.''
For simplicity, we do not do that. \qef
\end{remark}

If the channel variable $y$ is not used later in the program, then we need not name it.
Then, as syntactic sugar, the rule \eqref{as1} may be rendered as
\begin{equation}
 \mathtt{output}\ M(\bu) \label{as2}.
\end{equation}
If several \texttt{output} commands occur in the same program, they are presumed to refer to distinct variables $y$, so that they avoid write conflicts.

More generally, suppose that the gate $G$ is assigned potential measurements  $M_1, \dots, M_p$ where $p\ge1$.
The actual measurement, performed by $G$, depends on  the classical outcomes produced by the classical sources of $G$.
Let \F\ be the set of these classical sources,
and let $\bx$ be the indexed set \iset{x_F: F\in\F} of the classical outcomes produced by the classical sources%
\footnote{If you fix a linear order on the gates, then the indexed set \iset{x_F: F\in\F} could be replaced by an ordered tuple $(x_{F_1}, x_{F_2}, \dots)$ where $F_1 < F_2 < \dots$\,.}.

In \S\ref{sb:qc} we mentioned a choice function $\sigma_G$ that, given \bx, determines the actual measurement $\sigma_G(\bx)$.
While choice functions are convenient for the purposes of analysis, they are less convenient for programming purposes.
Instead we use Boolean guards $\beta_1(\bx), \dots, \beta_p(\bx)$ such that, if $\beta_q(\bx)$ holds and $q$ is the smallest index\footnotemark\ for which $\beta_q(\bx)$ holds, then
$M_q(\bu)$ is executed, and, if none of the guards holds, then $M_{p+1}(\bu)$ is executed.
(Notice that the choice function has a finite domain and a finite range of possible values, and so the desired guards always exist.)
Accordingly, instead of \eqref{as1}, we have a conditional assignment
\begin{equation}\label{as3}
\begin{aligned}
& \If     &\beta_1(\bx)\quad &\Then y := M_1(\bu) \\
& \Elseif &\beta_2(\bx)\quad &\Then y := M_2(\bu) \\
& \vdots \\
& \Elseif &\beta_p(\bx)\quad &\Then y := M_p(\bu)\\
& \Else &              &\phantom{\Then} y := M_{p+1}(\bu),
\end{aligned}
\end{equation}%
where $y$ is as in \eqref{as1}.
\footnotetext{Typically, the guards are disjoint, in which case ``smallest index'' amounts to ``unique index.''}
Generalizing \eqref{as2}, if the variable $y$ is not used later in the program, then, as syntactic sugar, the assignments $y := M_i(\bu)$ can be rendered as
$\mathtt{output}\ M_i(\bu)$ respectively.

In common examples in the literature, if a gate $G$ is assigned $p+1$ potential measurements, then $p=1$ and $M_2$ is the identity measurement.
In these common situations, \eqref{as3} simplifies to
\[
 \If\ \beta_1(\bx)\quad \Then y := M_1(\bu).
\]

\subsection{Bouts}
\label{bout}

If a bout $B$ is the set \set{G_j: j\in J} of gates $G_j$ and if rule $R_j$ programs $G_j$, then the \emph{parallel composition rule}
\begin{equation}\label{par1}
 \texttt{forall $j$ in $J$:  $R_j$ }
\end{equation}
programs $B$.
If ambiguity arises about where exactly the rule \eqref{par1} ends, we may use braces around the rule:
\begin{equation*}
\texttt{\{forall $j$ in $J$:  $R_j$\}}.
\end{equation*}

In the common case where $J$ consists of consecutive integers from $a$ to $b$, we write $[a,b]$ as an alternative notation for $J$, so that \eqref{par1} becomes
\begin{equation}
 \texttt{forall $j$ in $[a,b]$:  $R_j$ }.
\end{equation}

If $J = \set{j_1, j_2, \dots, j_k}$, then the parallel composition may also be rendered as
\begin{equation}\label{par2}
 R_{j_1} \parallel R_{j_2} \parallel \cdots \parallel R_{j_k}\:.
\end{equation}
The order of the components in this parallel composition can be chosen arbitrarily.
According to the semantics in \cite[\S6]{G141}, all these choices behave the same.

\subsection{Programs}
\label{sb:prog}

Let $A$ be a QC algorithm given by a circuit $C$ and schedule $B_1,\dots,B_t$ and suppose that the bouts $B_s$ are programmed by rules $R_s$.
A program for our algorithm may be written as a \emph{sequential composition}
\begin{equation}\label{seq1}
R_1; R_2; \dots; R_t
\end{equation}
which has an alternative form
\begin{equation}\label{seq2}
 \texttt{for $j=1$ to $t$:\  $R(j)$}
\end{equation}
where $R(j)$ is a rule such that $R(1)$ is equivalent to $R_1$, $R(2)$ is equivalent to $R_2$, etc.
In applications where such $R(j)$ is natural, the alternative form is convenient; see \S\ref{e:qft} for example. But such $R(j)$ can always be constructed artificially:
\begin{equation}\label{seq3}
\begin{aligned}
&\If\     &&j=1\   &\Then\quad &R_1  \\
&\Elseif\ &&j=2\   &\Then\quad &R_2  \\
&\vdots \\
&\Elseif\   &&j=t  &\Then\quad &R_t \\
&\Else\   &&       &\phantom{\Then}\quad &\texttt{Skip}
\end{aligned}
\end{equation}
where the rule \texttt{Skip} does nothing (and the \Else\ clause is included only to comply with the syntax in \S\ref{s:langb}).

As before, if ambiguity arises about where exactly the rule \eqref{seq2} ends, we may use braces around the rule:
\begin{equation*}
 \texttt{\{for $j=1$ to $n$:\ $R(j)$\}}\,.
\end{equation*}
The \texttt{for} construct is somewhat similar to the \texttt{forall} construct. In particular, in \eqref{par1} and in \eqref{seq2}, the variable $j$ is bound.
But notice that \texttt{for} operates sequentially, while
\texttt{forall} operates in parallel.
Also, the range of the variable $j$ in \texttt{forall} can be any finite set, but in \texttt{for} it has to be a segment of integers, not necessarily starting at 1 as in the examples above.

\begin{remark}[Normal form]
The program is a sequential composition of parallel compositions of assignments or conditional assignments.
Of course, we are not enforcing this normal form as a general requirement on QC-ASM programs.
\qef
\end{remark}

\subsection{Quantum input}\label{sb:qinput}

Let QC algorithm $A$ be as in \S\ref{sb:prog}.
The program \eqref{seq1} of $A$ says nothing about the initial state.

But a quantum circuit diagram for $A$ usually describes the initial state.
The set of wires of the circuit $C$ is partitioned into subsets $S_1, \dots, S_k$ of consecutive wires usually indicated by braces in the beginning of the diagram.
For our purposes, we drop the consecutivity requirement, so that $S_1, \dots, S_k$ can be an arbitrary partition of the set of wires.
Each $S_i$ is given by an ordered list $L_i$ of the members of $S_i$.
(In typical diagrams, that order is given by the top-to-bottom ordering of the wires.)
The diagram indicates the initial state \ket{\psi_i} on wires $L_i$, so that the intended initial quantum state of $A$ is (up to isomorphism) the tensor product
\begin{equation}\label{state1}
\ket{\psi_1}\ox\dots\ox\ket{\psi_k}.
\end{equation}

To match the diagrams, we introduce \emph{quantum input declarations}.
Recall the sort \qstate\ mentioned in the beginning of this section.
Expressions of sort \qstate\ denote $m$-qubit states for various values of $m$.
These  could be standard states, like one-qubit quantum states \ket0, \ket1 or two-qubit Bell states.
They could also be user-provided states.

The \emph{basic quantum input declaration} has the form
\begin{equation}\label{input1}
 \ket\psi\ \on\ L
\end{equation}
where \ket\psi\ is an $m$-qubit expression of type \qstate\ and $L$ is a list of $m$ distinct wires.

If quantum input declarations $D_1$ and $D_2$ deal with disjoint sets of wires, then
\[ D_1\ \et\ D_2 \]
is a quantum input declaration.
Similarly, we can combine any number of input declarations as long as their sets of wires are pairwise disjoint.
Suppose that sets $S_i$, lists $L_i$ and quantum states $\ket{\psi_i}$ are as above.
Then the initial state \eqref{state1} is described by the quantum input declaration
\begin{equation}\label{input2}
\ket{\psi_1}\ \on\ L_1\ \et\ \ket{\psi_2}\ \on\ L_2\ \et\ \dots\ \et\
         \ket{\psi_k}\ \on\ L_k\,.
\end{equation}

Sometimes, the rule \eqref{input2} or portions of it may be rendered more succinctly by means of the \texttt{forall} construct. For example, if every wire in a set $J$ is initialized with the same one-qubit state \ket\psi, we may write
\begin{equation}\label{input3}
\texttt{forall $j$ in $J$: \ket\psi\ on $j$}.
\end{equation}

We stipulate that the program $P$ of an algorithm may be preceded by a quantum input declaration $D$:
\begin{equation}
D; P.
\end{equation}

\begin{remark}[Quantum input rules]
Input declarations can be understood as input rules.
Think of \eqref{input1} as an instruction to input \ket\psi\ on wires $L$.
In \eqref{input2}, view \et\ as another notation for (or replace it by the sign \pr\ of) parallel composition.
\end{remark}

\subsection{Some perspective}
\label{sb:persp}

What are special features of the QC-ASM language from the perspective of the ASM theory and applications?

\begin{itemize}[leftmargin=0pt]
\item[]$\bullet\;$ The sort \texttt{measurement}.
\hspace{2pt}As far as we know, this sort does not appear in the ASM literature, with a possible exception of article \cite{GN} where the vocabulary is not explicit.
But this is isn't surprising.
It is common for an application area to have area-specific sorts.

\item[]$\bullet\;$ The invisible quantum state. \hspace{2pt}%
While the classical outcome is recorded in \eqref{as1} and \eqref{as3}, the quantum state is not mentioned in \S\ref{sb:gate}.
The reason is not neglect but the level of abstraction which,
as we mentioned above, is close to that of common circuit diagrams, e.g., those in \S\ref{s:examples}.
In classical algorithms, the actions of the operating system are typically below the algorithm's level of abstraction and so they are implicit.
In the case of QC algorithms, nature works in the background, like an operating system.
We revisit this issue in Remark~\ref{r:separate} in \S\ref{s:intrastep}.

\item[]$\bullet\;$ Sequential composition.
\hspace{2pt}There is no sequential composition construct in \cite{G141}; a program describes one generic step to be repeated over and over, possibly forever.
With this semantics in mind, the program \eqref{seq1} can be replaced with a version of \eqref{seq3} where each of $R_1, \dots, R_t$ increments $j$.
For practical purposes though, explicit sequential composition is convenient, and it appears already in \cite[\S3.1.2]{G103}.
It is used routinely in applications; see \cite{BS,G169} for example.

\item[]$\bullet\;$ Intra-step interaction with the environment.
\hspace{2pt}Consider the assignment $y := M(\bu)$ for example.
The right side is a call to the environment to perform the measurement $M(\bu)$.
Here $M$ is an \emph{external function call} in terms of ASM theory.
While external functions are not mentioned in \S6 of \cite{G141}, they make a modest debut in \S8.3 of that paper.
In applications, intra-step interaction is common; see \cite{BS,G169} for example.
Theoretically, intra-step interaction is formalized and analyzed in
\cite{G166-171,G176&182}.

\item[]$\bullet\;$ Calls to Nature.
\hspace{2pt}This feature is distinctive.
In a sense, this is not surprising.
With the exception of \cite{GN}, the ASM literature is devoted to classical (rather than quantum) algorithms and algorithms are usually described on a level of abstraction higher than the physical implementation where nature is directly involved.
We return to the issue in \S\ref{s:intrastep}.
\end{itemize}

\section{Examples}\label{s:examples}

For a few well-known QC algorithms, we give quantum circuit diagrams and QC-ASM programs.
We adopt ``frequently used quantum gates and circuit symbols'' listed in \cite[pages~xxx--xxxi]{NC}; specifically, we use the symbols $H, X, Z$, \swap, and \cnot\ (CNOT).
We use the notion of controlled version $cU$ of an operator $U$, as explained in \cite[\S4.3]{NC}.
It is presumed that the first qubit is the controller.
So, in $cU(v_0,v_1,\dots,v_k)$, the qubit on wire $v_0$ is the controller.

The one-qubit \emph{standard measurement} \SM\ produces 0 on qubit \ket0 and 1 on qubit \ket1 in the computational basis.
It is the projective measurement
\Iset{\ketbra ii: i\in\set{0,1}}.
The two-qubit \emph{parity measurement} \PM\ produces 0 on $\ket{00} = \ket0\ox\ket0$ and on \ket{11}, and produces 1 on \ket{01} and \ket{10} in the computational basis. In more detail, \PM\ is a projective measurement with classical outcomes 0,1 and the corresponding eigenspaces spanned by
$ \{\ket{00}, \ket{11}\} \text{ and } \{\ket{01}, \ket{10}\} $,
respectively.
According to our convention at the end of \S\ref{sb:expr}, $\SM(u)$ is the measurement \SM\ on the wire $u$, and $\PM(u,v)$ is the measurement \PM\ on wires $u,v$.

\begin{example}[Controlled-NOT]\label{e:cnot}
The two-qubit CNOT gate ﬂips the state of a target qubit if the control qubit is in the logical \ket1 state.
It is a single gate in most, but not all, approaches to quantum computing.
The following circuit, a version of the circuit in \cite{BDEK}, is a measurement-based implementation of the CNOT gate
that is useful for some optical realisations of quantum computing.
The circuit algorithm computes the CNOT operator on wires 1 (the controller) and 3 (the target).
The computation uses one ancillary qubit, on wire 2, which is eventually measured.

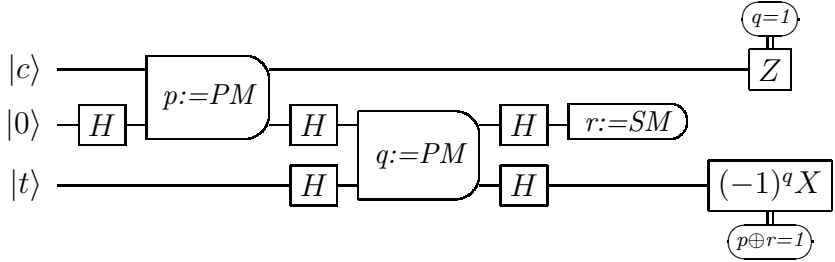
\hspace{-2pt}
\Qcircuit @C=8pt @R=.45em {
  &\ew &\ew &\ew &\ew &\ew &\ew
  &\emeasure{\textit{\scriptsize q=1}}\cwx[1] \\
\lstick{\ket{c}}
  &\qw &\multimeasureD{1}{\textit{\small p:=PM}} &\qw
  &\qw &\qw &\qw &\gate{Z} \\
\lstick{\ket0}
  &\gate{H} &\ghost{\textit{\small p:=PM}} &\gate{H}
  &\multimeasureD{1}{\textit{\small q:=PM}} &\gate{H}
  &\measureD{\textit{\small r:=SM}} \\
\lstick{\ket{t}}
  &\qw &\qw &\gate{H}
  &\ghost{\textit{\small q:=PM}} &\gate{H} &\qw &\gate{(-1)^q X} p\\
  &\ew &\ew &\ew &\ew &\ew &\ew
  &\emeasure{\textit{\scriptsize p\op r=1}}\cwx[-1]
}

Here is the corresponding QC-ASM program.
It is assumed that ``$\pr$'' binds more tightly than ``;''.

\begin{eatab}
\ \ket c on 1 \et\ \ket0 on 2 \et\ \ket t on 3;\\
\ $H(2);\ p:=\PM(1,2);\ H(2)\pr H(3);\ q:=\PM(2,3)$;\ $H(2)\pr H(3)$;
\  $r:=\SM(2)$;  \\
\ $\big(\If\ q=1\ \Then Z(1)\big) \pr
   \big(\If\ p\op r=1\ \Then (-1)^q X(3)\big)$
\end{eatab}

Taking into account that our setup is somewhat different from that in the literature \cite{BDEK,ZBL}, we check that the program transforms any of the four state vectors \ket{c0t} in the computational basis into \ket{cr(c\op t)}.
With four input vectors and three two-outcome measurements, there are $2^7 = 128$ different computation runs to track.
Symbolic verification reduces this task to tracking one generic computation run.


\begin{enumerate}
\item
Execute $H(2)$ in the initial  state \ket{c0t}. The new  state is
\[ \ket{c}\frac{\ket0+\ket1}{\sqrt2}\ket{t}  =
\frac1{\sqrt2}\ket{c}(\ket0+\ket1)\ket{t} =
\frac1{\sqrt2} \ket c\ket{c\op0}\ket t
  +\frac1{\sqrt2} \ket c\ket{c\op1}\ket t. \]

\item
Execute \Set{p := \PM(1,2)}.
If $p=0$ then the  state is projected into the eigenspace spanned by \set{\ket{00},\ket{11}} where \ket c\ket{c\op0} belongs, and if $p=1$ then the  state is projected into the eigenspace spanned by \set{\ket{01},\ket{10}} where \ket c\ket{c\op1} belongs.
The new state (after normalization) is
\[ \ket{c}\ket{c\oplus p}\ket{t}. \]

\item
Execute \Set{H(2)\| H(3)}. The new  state is
\begin{align*}
& \ket{c} \cdot
 \frac{\ket0+(-1)^{c+p}\ket1}{\sqrt2} \cdot
 \frac{\ket0+(-1)^t\ket1\ket1}{\sqrt2} = \\
& \frac12\ket{c}\big(\ket{00}+(-1)^{c+p+t}\ket{11}\big)+
  \frac12\ket{c}\big((-1)^t\ket{01}+(-1)^{c+p}\ket{10}\big)
\end{align*}

\item
Execute \Set{q := \PM(2,3)}.
If $q=0$ then the  state is projected into the eigenspace spanned by \set{\ket{00},\ket{11}} where the first summand belongs, and if $q=1$ then the  state is projected into the eigenspace spanned by \set{\ket{01},\ket{10}} where the second summand belongs.
Note that $qt$ is 0 or $t$ depending on whether $q$ is 0 or 1.
The new state is
\begin{align*}
& \frac1{\sqrt2} \ket{c} \cdot \left[(-1)^{qt}\ket0\ket{q} +
                 (-1)^{c+p+t+qt}\ket1\ket{q\op1}\right] = \\
& \frac1{\sqrt2} (-1)^{qt}\ket{c}\ket0\ket{q} +
  \frac1{\sqrt2} (-1)^{c+p+t+qt}\ket c\ket1\ket{q\op1}
\end{align*}

\item
Execute \Set{H(2)\| H(3)}. The new  state is
\begin{align*}
& \frac1{\sqrt2}(-1)^{qt}\ket c \cdot
  \frac{\ket{0}+\ket{1}}{\sqrt2} \cdot
  \frac{\ket{0}+(-1)^{q}\ket{1}}{\sqrt2} \quad \\
& +\quad \frac1{\sqrt2}(-1)^{c+p+t+qt}\ket c \cdot
\frac{\ket{0}-\ket{1}}{\sqrt2} \cdot
\frac{\ket{0} - (-1)^q\ket{1}}{\sqrt2}
\end{align*}

\item
Execute $r:=\SM(2)$. The new  state is
\begin{align*}
& \frac12(-1)^{qt}\ket{cr}\Big(\ket{0}+(-1)^{q}\ket{1}\Big)
+\ \frac12(-1)^{c+p+t+qt+r}\ket{cr}\Big(\ket{0}-(-1)^q\ket{1}\Big) =\\
&\begin{cases}
 (-1)^{qt}\ket{cr0} &\text{if }c\op t\op p\op r=0\\
 (-1)^{qt+q}\ket{cr1} &\text{if }c\op t\op p\op r=1
 \end{cases}\ =\
  (-1)^{q(c + p + r)}\ket{cr(c\op t\op p\op r)}
\end{align*}

\item
Execute
\Set{\big(\If\ q=1\ \Then Z(1)\big) \pr
   \big(\If\ p\op r=1\ \Then (-1)^q X(3)\big)}.
The final  state is the desired \ket{cr(c\op t)}.
\qedhere
\end{enumerate}
\end{example}

\begin{remark}[Parallelization]\label{r:par}
There is a more liberal view on the parallelism in quantum circuit diagrams.
For example, the first $H$ gate on wire~3  in the diagram above should be executed in parallel with the three-gate sequence preceding the second measurement, not necessarily after the first measurement.
Elsewhere we will discuss the liberal view and the QC-ASM reflection of it.
Here we just illustrate a version of that reflection on the example of the above diagram:

\begin{eatab}
\ \ket c on 1 \et\ \ket0 on 2 \et\ \ket t on 3;\\[3pt]
\ $\Big(H(2);\ p:=\PM(1,2);\ H(2)\Big)\pr H(3);\ q:=\PM(2,3)$; \\
\ $\big(\If\ q=1\ \Then Z(1)\big) \pr
 \big(H(2); r:=\SM(2)\big) \pr H(3)$; \\
\ $\big(\If\ p\op r=1\ \Then (-1)^q X(3)\big)$
\end{eatab}
\end{remark}

\begin{example}[Teleportation]\label{teleport}
The following diagram is (a version of) the teleportation diagram in Figure~1.13 in \cite{NC}.

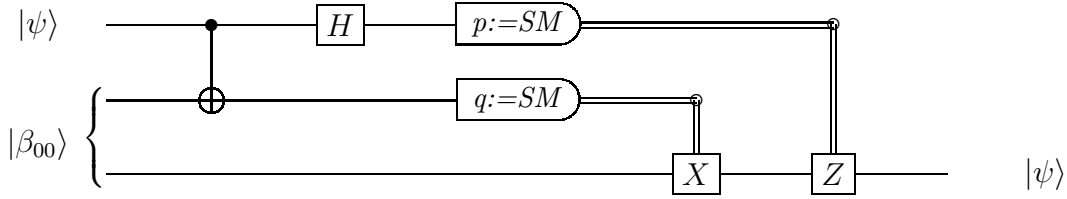
\[\phantom{mm} \Qcircuit @C=35pt @R=12pt {
\lstick{\ket\psi\quad}
 &\ctrl{1} &\gate{H} &\measureD{\textit{\small p:=SM}}
 &\cw &\circ\cw\cwx[2]\\
\lstick{}
 &\targ   &\qw &\measureD{\textit{\small q:=SM}}
 &\circ\cw\cwx[1]\\
\lstick{}
 &\qw &\qw &\qw  &\gate{X}  &\gate{Z} &\qw
 &\rstick{\hspace{-1em}\ket\psi}
\inputgroupv{2}{3}{0.8em}{1.4em}{\ket{\beta_{00}}\quad}
}\]
where \ket{\beta_{00}} is a Bell state.
\noindent
The corresponding QC-ASM program is

\begin{eatab}
\> \ket\psi\ on 1\ \et\ \ket{\beta_{00}} on 2,3;\\[3pt]
\> $CNOT(1,2);\ H(1);\ p:=\SM(1)\pr q:=\SM(2)$;\\
\> if $q=1$ then $X(3)$; if $p=1$ then $Z(3)$
\end{eatab}
\end{example}

\begin{example}[Quantum Fourier transform]\label{e:qft}\mbox{}
The QC algorithm in question computes a unitary operator known as the $n$-qubit quantum Fourier transform $\QFT_n$.
The algorithm is described in \cite[\S5.1]{NC}.
The following diagram is Figure~5.1 in \cite{NC}, except that we don't show the output state (because it isn't a part of the algorithm).
The diagram depicts the essential part of the algorithm.
The ellipses refer to omitted details.
Also as in \cite[\S5.1]{NC}, ``not shown are swap gates at the end of the circuit which reverse the order of qubits.''

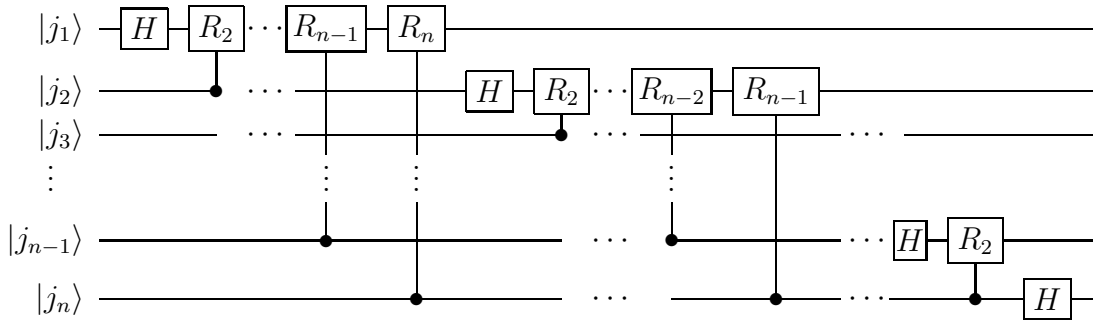
\[
\Qcircuit @C=8pt @R=6pt {
&\lstick{\ket{j_1}} 
 &\gate{H} &\gate{R_2} &\cdots &\push{\gt{$R_{n-1}$}}\qwx[3] &\gate{R_n}\qwx[3]
 &\qw &\qw &\qw &\qw &\qw &\qw &\qw &\qw &\qw &\qw &\qw \\
&\lstick{\ket{j_2}} 
 &\qw &\ctrl{-1} &\cdots &\push{\rule{-2em}{0em}} &\qw
 &\gate{H} &\gate{R_2} &\cdots  &\push{\gt{$R_{n-2}$}}\qwx[2]
 &\gate{R_{n-1}}  &\qw &\qw &\qw &\qw &\qw &\qw\\
&\lstick{\ket{j_3}} 
 &\qw &\qw &\cdots  &\push{\rule{-2em}{0em}} &\qw
 &\qw &\ctrl{-1} &\cdots &\push{\rule{-2em}{0em}} &\qw &\qw
 &{\ \cdots} & &\qw &\qw &\qw\\
&\lstick{} 
 &&& &&& &&& &&& &\\
\mbox{\phantom{m}}\vdots\phantom{mmm} 
 &&&&&\vdots &\vdots  &&&&\vdots &&&& \\
&\lstick{} 
 &&& &&& &&& &&& \\
\lstick{} 
 &&& &&& &&& &&& &\\
&\lstick{\ket{j_{n-1}}} 
 &\qw &\qw &\push{\rule{-50pt}{0em}} &\ctrl{-1}
 &\qw &\qw &\qw &\cdots &\control\qwx{-1} &\qw &\qw
 &{\ \cdots} &\push{\ \gt{$H$}} &\gate{R_2} &\qw &\qw\\
&\lstick{\ket{j_n}} 
 &\qw &\qw &\qw &\push{\rule{-50pt}{0em}} &\ctrl{-2}
 &\qw &\qw &\cdots &  &\ctrl{-7} &\qw
 &{\ \cdots} &\push{\rule{-1em}{0em}} &\ctrl{-1} &\gate{H} &\qw}
\]
where $R_k = \begin{pmatrix}1&0\\0&e^{2\pi i/2^k}\end{pmatrix}$ in the computational basis.
The corresponding QC-ASM program is

\begin{eatab}
\> forall $i$ in $[1,n]$: \ket{j_i} on $i$;\\
\> \{for $i=1$ to $n$: \\
\>\> $H(i)$;\
   for $k=i+1$ to $n$:  $cR_{k-i+1}(k,i)$\};\\
\> forall $i=1$ in $\big[1,\lfloor n/2\rfloor\big]$:\  swap$(i,n-i+1)$
\end{eatab}
\end{example}

\medskip
\begin{example}[Quantum phase estimation]
The QC algorithm in question is described in \cite[\S5.2]{NC}.
The following diagram is an obvious completion of the diagram in
Figure~5.2 in \cite{NC}; in particular $U$ is the same unitary operator as there.
The diagram depicts the algorithm.
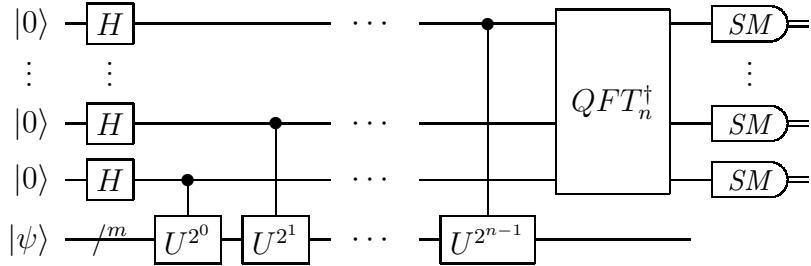
\[
\Qcircuit @C=8pt @R=6pt {
\lstick{\ket0}
 &\gate{H} &\qw &\qw &\qw &\push{\cdots\quad} &\ctrl{4}
  &\multigate{3}{\QFT_n\dg} &\qw &\measureD{\textit{\small SM}} &\cw\\
{\Vdots}\phantom{mmn}
 &\Vdots & & & & & &
  \nghost{\QFT_n\dg} & &\Vdots\\
\lstick{\hbox{\ket0}}
 &\gate{H} &\qw &\ctrl{2} &\qw &\push{\cdots\quad} &\qw
  &\ghost{\QFT_n\dg} &\qw &\measureD{\textit{\small SM}} &\cw\\
\lstick{\ket0}
 &\gate{H} &\ctrl{1} &\qw &\qw &\push{\cdots\quad} &\qw
  &\ghost{QFT_n\dg} &\qw &\measureD{\textit{\small SM}} &\cw\\
\lstick{\ket\psi}
 &/^m \qw &\gate{U^{2^0}} &\gate{U^{2^1}} &\qw &\push{\cdots\quad}
  &\gate{U^{2^{n-1}}} &\qw  &\qw \qw
}
\]
Here is the corresponding QC-ASM program:

\begin{eatab}
\> \{forall $i$ in $[1,n]$: \ket0 on $i$\}\ \et\ \ket\psi\ on  $n+1,\dots,n+m$;\\
\> forall $i$ in $[1,n]$:  $H(i)$;\\
\> for $i=1$ to $n$:
 $c U^{2^{i-1}}(n-i+1,(n+1),\dots,(n+m))$;\\
\> $\QFT\,\dg_n(1,\dots,n)$;\\
\> forall $i$ in $[1,n]$: output $\SM(i)$
\end{eatab}
\end{example}

\medskip
\begin{example}[Grover's algorithm]\label{grover}
Grover's search algorithm \cite{Grover} is one of the most famous
quantum algorithms.
In \cite{G240}, building on the exposition of Grover's algorithm in \cite{Mermin}, we attempted to explain the algorithm to non-physicists.
The following diagram is a variation on the diagram \cite{Wiki}.
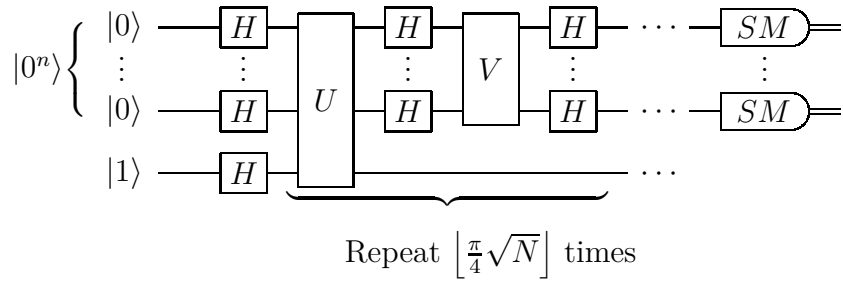
\[\phantom{mmmm}\Qcircuit @C=1em @R=.7em {
\lstick{\ket{0}} & \qw & \gate{H} & \multigate{3}{U} & \gate{H} & \multigate{2}{V} & \gate{H} & \qw & \cdots & & \measureD{\SM} & \cw \\
\Vdots\phantom{mmn} & & \Vdots & & \Vdots & & \Vdots & & & & \Vdots \\
\lstick{\ket{0}} & \qw & \gate{H} & \ghost{U} & \gate{H} & \ghost{V} & \gate{H} & \qw & \cdots & & \measureD{\SM} & \cw \\
\lstick{\ket{1}} & \qw & \gate{H} & \ghost{U} & \qw & \qw & \qw & \qw & \cdots & \\
&&&&& \dstick{\text{Repeat %
$\big\lfloor\frac{\pi}{4}\sqrt{N}\big\rfloor$ times}}
\gategroup{1}{4}{4}{7}{.7em}{_\}}
\inputgrouph{1}{4}{1.35em}{\ket{0^n}\Bigg\{}{3.5em}
}\]

\bigskip\bigskip\noindent
Here $U$ is a quantum search oracle.
For our purposes, it is not important what exactly $U$ is;
it is just a given unitary operator.
$V$ is the unitary $2\ketbra{0^n}{0^n}-\Id_n$.
The corresponding QC-ASM program is as follows.

\begin{eatab}
\> \{forall $i$ in $1\dots n$: \ket0 on $i$\}\ \et\ \ket1 on $n+1$;\\
\> forall $i$ in $1\dots(n+1)$:  $H(i)$; \\
\> \{for $k=1$ to $\big\lfloor\frac\pi4\sqrt N\big\rfloor$: \\
\>\>\ $U(1,\dots,n+1)$;\
\>\>\>\>\>\  forall $i$ in $1\dots n$:  $H(i)$;\\
\>\>\ $V(1,\dots,n)$;
\>\>\>\>\>\  forall $i$ in $1\dots n$:  $H(i)$  \\
\>\ \};\\
\> forall $i$ in $1\dots n$:  output $\SM(i)$
\end{eatab}
\end{example}

\section{QC algorithms as  interactive algorithms}
\label{s:intrastep}

Article \cite{G141} axiomatized \emph{sequential algorithms}, defined \emph{sequential abstract state machines}, in short \emph{sequential ASMs}, and proved that every sequential algorithm is step-for-step simulated by an appropriate sequential ASM. (Sequential ASMs are sequential algorithms of course.)

A sequential algorithm does not explicitly interact with the environment during a step.
More exactly, it does not interact with the environment on its abstraction level, in the following sense.
The environment may intervene implicitly: computing ``given'' functions,  implementing the updates produced by the algorithm, etc.
Such interventions are not treated as actual interaction and viewed as being below the abstraction level of the algorithms.

QC algorithms rely on the environment to perform measurements, and thus they involve explicit intra-step interaction with the environment.

\begin{remark}\label{r:separate}
Performing a measurement involves evaluating the classical outcome and applying the corresponding linear operator to update the quantum state.
It is tempting to treat the classical outcome as a genuine choice of the environment and treat the update of the quantum state as an action of the algorithm (implemented by the environment).
Such separation makes mathematical sense and is similar to common practice in classical computing.
For example, the conditional assignment
\begin{equation*}
\begin{aligned}
& \If\ \random=0 &\Then &x := 7 \\
& \Else         &      &x := 11,
\end{aligned}
\end{equation*}
where \random\ is a bit chosen at random by the environment (operating system), could be performed entirely by the environment.
But normally \random\ would be chosen by the environment, and the update of $x$ would be performed by the algorithm.
However, the separation is thought to have little physical sense for QC algorithms.
When nature evaluates the classical outcome, it simultaneously updates the quantum state.
Also quantum circuit diagrams show no separation.
Neither will we. \qef
\end{remark}

To account for intra-step interaction with the environment, we generalized in \cite{G166-171} the theory of sequential algorithms, developed in \cite{G141}, to the theory of so called \emph{ordinary interactive small-step algorithms}, in short \emph{interactive algorithms}.
An interactive algorithm
\begin{itemize}
\item proceeds in discrete steps,
\item performs only a bounded amount of work in each step,
\item uses only such information from the environment as can be regarded as answers to the queries issued by the algorithm, and
\item never completes a step until all queries issued in that step have been answered.
\end{itemize}

In Part~I of \cite{G166-171}, we axiomatized  interactive algorithms.
In Part~II, we defined  interactive abstract state machines, in short \emph{ interactive ASMs}.
In Part~III, we proved that every  interactive algorithm can be step-for-step simulated by an  interactive ASM.

QC algorithms are  interactive algorithms.
Think of $M(\bu)$ in \eqref{as1} as a function call to the environment.
Of course the environment needs access to the appropriate quantum state \ket\psi.
But, staying on the abstraction level of quantum circuit diagrams, we deal with the quantum state implicitly.

In terms of  interactive algorithms, the measurement, a tuple of the relevant wires, and the relevant quantum state form a query.
The answer comprises the classical outcome of $M(\bu)$ and the updated quantum state.

QC algorithms satisfy a constraint
---namely, no query depends on any answer from the same step ---
that dramatically simplifies the theory developed in \cite{G166-171}.

\begin{remark}[Simplification]\label{r:simple}
Readers who want to study \cite{G166-171} as background for the present paper would do well to keep this simplification in mind.
To illustrate the simplification, we note that in the simplified version, we have the following.
\begin{itemize}
\item In the Interaction Postulate, the causality relation between answer functions and potential queries reduces to just the set of potential queries (caused by the empty answer function).
\item The notion of context (Definition~5.5 in Part~I) becomes nearly trivial. A context just assigns an answer to each potential query. \qef
\end{itemize}
\end{remark}

In fact, a QC algorithm given by a program of \S\ref{s:lang}
is an interactive abstract state machine \cite[Part~II]{G166-171}.
In the terminology of that paper, in the assignment $y:= M(\bu)$, $M$ is a so-called \emph{external} function symbol, and $M(\bu)$ is a \emph{query} to the environment.
The environment evaluates the query and returns the result to the algorithm.
The case of conditional assignment \eqref{as3} is similar;
$M_1, \dots, M_p, M_{p+1}$ are external function symbols there.

In the current situation, with QC algorithms, evaluating the query means performing the measurement, and the result returned to the algorithm is the classical outcome.
In addition, the environment updates the quantum state.
Since we agreed to treat the quantum state implicitly, the algorithm does not ``see'' the quantum state; the update of the quantum state is exclusively the business of the environment.
This may seem delegating too much to the environment;
in this connection, see Remark~\ref{r:separate}.

Still, the theory of  interactive algorithms is more involved than that of sequential algorithms.
By slightly lowering the level of abstraction of quantum circuit diagrams, we can turn intra-step interaction of QC algorithms with their environments into inter-step interaction, so that QC algorithms become sequential algorithms.
Taking this into account, we develop, in the next section, an alternative approach to QC algorithms, using only inter-step interaction, so that QC algorithms are sequential.

\section{QC algorithms as sequential algorithms}
\label{s:interstep}

The programming language of \S\ref{s:lang} is on the desired level of abstraction, that of quantum circuit diagrams.
The purpose of the present section is not to change that language, but rather to show that a slight refinement of the language provides the proper semantics without intra-step interaction.

Assignment \eqref{as1} is simulated by a two-step procedure of the algorithm in collaboration with the environment that, between the two steps of the algorithm, evaluates the measurement at the appropriate quantum state, updates the quantum state, and returns the classical outcome.

The algorithm starts by preparing the environmental intervention.
In terms of  interactive algorithms, it composes a query.
The measurement $M$ (the indexed set of operators) to be evaluated is assigned to a special variable mq (an allusion to ``measurement in the query''), and the tuple $\bu$ of relevant wires is assigned to a special variable wq (an allusion to ``wires in the query'').
The environment would also need the state \ket\psi\ of the quantum system.
A special variable could be introduced for that purpose, but we chose
to keep the quantum state implicit, as it is usually done on quantum
circuit diagrams and as it is in \S\ref{s:lang}.

Then the environment evaluates the measurement mq in the quantum state \ket\psi, updates the quantum state, and stores the classical outcome in a variable $o(\wq)$.
(The purpose of the argument $\wq=\bu$ is to disambiguate the storage location. There could be several queries that the environment evaluates in parallel, even several uses of the same measurement on different wires.)
Finally, the algorithm assigns the classical outcome to a fresh variable, called $y$ in \eqref{as1}.
Here is that procedure:
\begin{equation*}
\begin{aligned}
&\mq := M\ \parallel\ \wq:=\bu; \\
&y := o(\bu)
\end{aligned}
\end{equation*}%
where, in between the two steps, the environment performs the measurement \mq(\wq) at the quantum state, updating the state in the process, and returns the classical outcome as $o(\wq)$.

Similarly, conditional assignment \eqref{as3} is simulated by a two-step procedure of the algorithm together with one step of the environment sandwiched between the two steps of the algorithm:
\begin{equation*}
\begin{aligned}
& \If\ \    \beta_1(\bx) &\Then \mq := M_1\\
& \Elseif\ \ \beta_2(\bx) &\Then \mq := M_2\\
& \vdots \\
& \Elseif\ \ \beta_p(\bx) &\Then \mq := M_p\\
& \Else\                &      \mq := M_{p+1}\\
& \parallel\quad  \wq := \bu\,;\\[3pt]
& y := o(\bu)
\end{aligned}
\end{equation*}
Again, in between the two steps, the environment performs the measurement \mq(\wq) at the quantum state, updating the state in the process, and returns the classical outcome as $o(\wq)$.
And again, the algorithm stores that outcome as $y$.

\section{Syntax and semantics of QC-ASM rules}
\label{s:langb}

\subsection{Expressions}\label{sb:expr}

We defined vocabularies and expressions in  \S\ref{s:lpre}.
To describe QC algorithms as abstract state machines, we need some additional assumptions and terminology about vocabularies and expressions.

We define \emph{wire names} as expressions of type integer which have no programming variables and can be evaluated at compile time.
(Example~\ref{e:qft} uses wire names $1, 2, \dots, n$ where $n$ is not a programming variable but a parameter whose value is fixed ahead of time in each QFT circuit.)

\emph{Channel variables} form a subcategory of programming variables.

\paragraph{Sorts}
For each positive integer $k$, we have sorts \texttt{$k$-qubit unitary} and \texttt{$k$-qubit measurement}; we also have sort \texttt{qstate}.
These sorts are \emph{quantum sorts}; the other sorts are \emph{classical}.
We stipulate that channel variables must be of classical sorts.

As in \S\ref{sb:meas}, a $k$-qubit unitary $U$ may be viewed as a degenerate $k$-qubit measurement $M_U$ with a unique classical output and unique linear operator $U$; this view allows us to simplify the exposition.
Still, formally speaking, there is difference between a unitary $U$ and the measurement $M_U$.
This is similar to viewing an integer as a real number.

For brevity, we say that an expression $e$ is \emph{classical} if the values of $e$ and all its subexpressions are of classical sorts.

\subsection{Generalized circuits}

Recall that a gate $G$ is assigned a collection of potential measurements and that the selection function $\sigma_G$ selects, at run time, the actual measurement depending on the values coming to $G$ via the incoming classical channels.

For technical reasons, we need to generalize the notion of quantum circuit.
To motivate the desired generalization, consider the possibility of breaking a quantum circuit $C$ into an initial segment $C_1$ and the complementary final segment $C_2$.
Are those two segments genuine quantum circuits?
More precisely, suppose that $C_1$ is a proper initial segment (initial in the sense that it is closed under prerequisites) of the quantum circuit $C$, and let $C_2$ be the complementary final segment.
If an exit of a $C_1$ gate is bound to an entry of a $C_2$ gate, bind that gate exit to a fresh output node in $C_1$, and bind a fresh input node to that gate entry in $C_2$.
As a result, $C_1$ is a quantum circuit in its own right, an initial subcircuit of $C$.
If there are no classical channels from $C_1$ gates to $C_2$ gates, then $C_2$ is also a quantum circuit.
Otherwise $C_2$ has incoming classical channels with sources outside $C_2$; it is in that sense that $C_2$ is a generalized quantum circuit.

If a $C_1$ gate $F$ is a classical source of one or more $C_2$ gates, think of the value that $F$ puts into its outgoing channel as the value of a special channel variable set by $F$.

\begin{definition}
Let \O\ be a set of channel variables.
(Intuitively, the values of \O\ variables are set by entities outside of the circuit under consideration.)
\emph{\O-circuits} are the generalization of quantum circuits where, for each gate $G$, the arguments of the selection functions $\sigma_G$ are not only the values sent by the classical sources of $G$ but also the values of \O\ variables. \qef
\end{definition}
\noindent
Quantum circuits are \O-circuits where $\O=\emptyset$.

\subsection{Syntax of rules}

Let \O\ be a finite set of channel variables.
We define the notion of \emph{rule over \O}, in short \emph{\O-rule}.

The definition is by induction.
In the process of induction, we also define, for every \O-rule $R$, the following.
\begin{itemize}[itemsep=3pt,topsep=3pt]
\item[-] The \emph{wire set} $W(R)$ of $R$.
The intention is that $W(R)$ is the set of wires involved in the execution of $R$.
\item[-] The set \Sub{R} of \emph{subrules} of $R$.
We stipulate that $R$ is a subrule of itself; the induction is used to define the \emph{proper} (different from $R$) subrules.
\item[-] The set \OV{R} of \emph{output variables} of $R$ which are channel variables occurring on the left side of assignments (that is assignment targets) in $R$.
\end{itemize}

\begin{definition}
\begin{enumerate}[itemsep=3pt,topsep=3pt]
\item \textit{Quantum assignments}. If $k,p$ are positive integers, $e_1, \dots, e_{p+1}$ are expressions of type \texttt{$k$-qubit measurement}, \bu\ is a tuple of wire names $(u_1, \dots, u_k)$ with distinct values, $\beta_1, \dots, \beta_p$ are boolean-valued expressions with no subexpressions of quantum sorts and whose only variables are channel variables in \O, and $y$ is a fresh channel variable, then the conditional assignment
\begin{equation}\label{as}
\begin{aligned}
& \If     &\beta_1\quad &\Then y := e_1(\bu) \\
& \Elseif &\beta_2\quad &\Then y := e_2(\bu) \\
& \vdots \\
& \Elseif &\beta_p\quad &\Then y := e_p(\bu) \\
& \Else   &   &\phantom{\Then} y := e_{p+1}(\bu)
\end{aligned}
\end{equation}
is an \O-rule, a \emph{gate \O-rule}, with wire set \set{u_1, \dots, u_k}, with no proper subrules, and with the single output variable $y$.

(On syntactic grounds, one may reasonably expect that assignments ``$y:=e_q(\bu)$'' are subrules of the gate \O-rule. But, semantically, they are not; they describe gates other than the one described by the gate \O-rule.)

We stipulate that a measurement expression may occur only within a gate rule, on the right side of an assignment.

\item \textit{Classical assignments}.
A (properly typed) assignment
\[ f(e_1,\dots,e_r) := e_0, \]
where both sides of the assignment are classical and where $f(e_1,\dots,e_r)$ isn't a channel variable,
is an \O-rule.
It has the empty sets of wires, of proper subrules, and of output variables.
Such a rule is called a \emph{classical assignment}.

\item \textit{Classical conditionals}.
If $n$ is a positive integer, if $R_1, \dots, R_{n+1}$ are \O-rules involving only classical expressions, and if $\beta_1, \dots, \beta_n$ are classical boolean-valued expressions, then the conditional
\[
\begin{aligned}
& \If     &\beta_1(\bx)\quad &\Then R_1 \\
& \Elseif &\beta_2(\bx)\quad &\Then R_2 \\
& \vdots \\
& \Elseif &\beta_n(\bx)\quad &\Then R_n,\\
& \Else   &                  &\phantom{\Then} R_{n+1}
\end{aligned}
\]
is an \O-rule.
It has the empty sets of wires, the set $\Sub{R_1} \cup \cdots \cup \Sub{R_{n+1}}$ of proper subrules, and the empty set of output variables.

\item \textit{Parallel compositions}.
If \O-rules \iset{R_j: j\in J} have pairwise disjoint wire sets and if no output variable of any constituent $R_j$ occurs in the other constituents, then
\[  \texttt{forall $j$ in $J$:  $R_j$ } \]
is an \O-rule with the wire set $\bigcup_j W(U_j)$, the set $\bigcup_j \Sub{R_j}$ of proper subrules, and the set $\bigcup_j \OV{R_j}$ of output variables.

\item \textit{Sequential compositions}.
Let $R_1, R_2$ be rules over sets $\O, \O'$ of channel variables respectively such that their sets of output variables are disjoint. Suppose that $\O' = \O \cup \OV{R_1}$. Then \[ R_1; R_2 \] is an $\O$-rule with the wire set $W(R_1) \cup W(R_2)$,  the set $\Sub{R_1} \cup \Sub{R_2}$ of proper subrules, and the set $\OV{R_1} \cup \OV{R_2}$ of output variables.

The sequential composition $R_1; R_2; \dots; R_n$ of many components is defined similarly. \qef
\end{enumerate}
\end{definition}

When we discuss semantics below, it will be clear that the multi-component sequential composition is equivalent to the result of composing the same rules two at a time.
Sequential composition is associative.

We say that an \O-rule $R$ is \emph{classical} if all expressions in $R$ are classical; otherwise $R$ is \emph{quantum}.

Below, by default, a rule is an \O-rule where \O\ is the empty set.

\subsection{Semantics of rules}

In \S\ref{sb:what}, we defined a schedule for a quantum circuit $C$ as a sequence of disjoint bouts such that every gate is in one of the bouts with the prerequisites in the earlier bouts.
A schedule  $B_1, B_2, \dots, B_t$ is uniquely characterized by the partial order
\[ F<G \iff \exists i,j (F\in B_i \land G\in B_j \land i<j) \]
that extends the prerequisite relation.
The characterization motivates a generalization of the notion of schedule that we will arrive at later in this section.

Each quantum \O-rule $R$ determines an \O-circuit $C(R)$ and a (strict) partial order $<_R$ for $C(R)$.
$C(R)$ and $<_R$ are defined by the following induction where
the set of gates of $C(R)$ will be denoted \G{R}.
\begin{itemize}
\item If $R$ is a quantum assignment \eqref{as}, then $C(R)$ comprises a single gate described by $R$. That gate acts on wires \bu, its measurements are given by expressions $e_q$, and its selection function chooses the least $q$ such that $\beta_q$ is true; if no $\beta_q$ is true, it chooses $e_{p+1}$.
    And $<_R$ is the empty partial order.
\item Let $R$ be the parallel composition of quantum \O-rules \iset{R_j: j\in J} and maybe some classical \O-rules.
    Then $C(R)$ is the obvious ``disjoint union'' of the circuits $C(R_j)$.
    (Recall that the wire sets $W(R_j)$ are pairwise disjoint.)
    Further, $<_R$ is the parallel composition of partial orders $<_{R_j}$.
    In set theoretic terms (that is, viewing partial orders as sets of ordered pairs), $<_R$ is the disjoint union of the partial orders $<_{R_j}$.
\item Suppose that $R$ is a sequential composition $R_1; R_2$ over sets $\O, \O'$ of channel variables respectively.
    If one of the two rules, $R_j$ is classical, then $C(R)$  is the circuit of the other rule, $C(R_{3-j})$, and $<_R$ is $<_{R_{3-j}}$.
    Suppose that both rules are quantum.
    The $\O$-circuit $C(R)$ is as follows.

    \G{R} is the disjoint union of \G{R_1} and \G{R_2}.
    The Bind relation of $C(R)$ is the least binary relation which includes the disjoint Bind relations of $C(R_1)$ and $C(R_2)$ and which contains the pairs $(a,b)$ such that, in $C(R_1)$, $a$ is a gate exit bound to the output node on some wire $u$, while, in $C(R_2)$, the input node on wire $u$ is bound to a gate entry $b$.
    The quantum source relation of $C(R)$ is determined by the Bind relation.
    The classical source relation of $C(R)$ is the disjoint union of the classical source relation of $C(R_1)$,  the classical source relation of $C(R_2)$, and the set of pairs $(F,G) \in \G{R_1} \times \G{R_2}$ such that the channel variable of $F$ occurs in the gate subrule of $R_2$ that defines $G$.

    Further, $<_R$ is the series composition of partial orders $<_{R_1}$ and $<_{R_2}$.
    In set theoretic terms, $<_R$ is the disjoint union of $<_{R_1}, <_{R_2}$, and $\G{R_1} \times \G{R_2}$. \qef
    \end{itemize}

Recall that \emph{series-parallel} partial orders constitute the smallest class of partial orders that contains the one-element partial orders and is closed under parallel and series compositions \cite[p.112]{Mohring}.
The partial schedules $<_R$ that arise in the semantics of rules $R$ are exactly the series-parallel partial orders.

\begin{definition}
\begin{itemize}
\item A \emph{partial schedule} for an \O-circuit $C$ is any series-parallel partial order $<$ on the set of $C$ gates that extends the prerequisite relation.
    A \emph{partial \O-algorithm} is given by an \O-circuit $C$ and a partial schedule $<$ for $C$.
\item Semantically, each \O-rule $R$ means the partial \O-algorithm $(C(R), <_R)$. \qef
\end{itemize}
\end{definition}

Every series-parallel partial order $P$ with more than one element is a (nontrivial) parallel composition or series composition (but not both).
Thus, each such $P$ can be represented (up to isomorphism) by a \emph{decomposition tree} \cite[p.~113]{Mohring}.
Moreover, each \O-rule $R$ determines a decomposition tree $T_R$ for $<_R$, essentially given by the parse tree of $R$.

Furthermore, any series-parallel partial order has a unique
\emph{canonical decomposition}\footnotemark\ tree
\cite[p.~113]{Mohring}.
\footnotetext{If $P$ is a series composition, keep decomposing it (as a series composition) until none of the components is a series composition. Do similarly for parallel decompositions.}
Of course, the canonical decomposition tree for $<_R$ need not coincide with $T_R$, even though both trees represent the same partial order.
One may want to call a rule $R$ \emph{canonical} if $T_R$ is a canonical decomposition tree.

\end{document}